# Sliding Ferroelectric Tunnel Junctions

Jie Yang,[1, 2] Jun Zhou,[3] Jing Lu,[1, 4, 5, 6] Zhaochu Luo,[1] Jinbo Yang,[1, 4, 5, 6, †] and Lei Shen[2,*]

[1] State Key Laboratory for Mesoscopic Physics and School of Physics, Peking University, Beijing 100871, P. R. China
[2] Department of Mechanical Engineering, National University of Singapore, 9 Engineering Drive 1, Singapore 117575, Singapore
[3] Institute of Materials Research & Engineering, A*STAR (Agency for Science, Technology and Research), 2 Fusionopolis Way, Innovis, Singapore 138634, Singapore
[4] Collaborative Innovation Center of Quantum Matter, Beijing 100871, P. R. China
[5] Beijing Key Laboratory for Magnetoelectric Materials and Devices (BKL-MEMD), Peking University, Beijing 100871, P. R. China
[6] Peking University Yangtze Delta Institute of Optoelectronics, Nantong 226010, P. R. China

Email: [†]jbyang@pku.edu.cn; [*]shenlei@nus.edu.sg

Very recently, ferroelectric polarization in staggered bilayer hexagonal boron nitride (BBN) and its novel sliding inversion mechanism was reported experimentally (*Science* **2021**, *372*, 1458; **2021**, *372*, 1462), which paves a new way to realize van der Waals (vdW) ferroelectric devices with new functionalities. Here, we develop vdW sliding ferroelectric tunnel junctions (FTJs) using the sliding ferroelectric BBN unit as ultrathin barriers and explore their transport properties with different ferroelectric states and metal contacts via the first principles. It is found that the electrode/BBN contact electric field quenches the ferroelectricity in the staggered BBN, resulting a very small tunnelling electroresistance (TER). Inserting high-mobility 2D materials between Au and BN can restore the BBN ferroelectricity, reaching a giant TER of ~10,000% in sliding FTJs. We finally investigate the metal-contact and thickness effect on the tunnelling property of sliding FTJs. The giant TER and multiple non-volatile resistance states in vdW sliding FTJs show the promising applications in voltage-controlled nano-memories with ultrahigh storage density.

**Introduction**

Since the first report of ferroelectricity in 1921,[1] this phenomenon has attracted tremendous interest in fundamental physics and practical device applications. All ferroelectrics discovered are polar in bulk, including the recently reported two-dimensional (2D) ferroelectric materials whose 3D counterparts are polar materials.[2-7] Recently, a whole new type of ferroelectrics was theoretically predicted first,[8-9] and then experimentally reported by parallel-stacking two layers of van der Waals (vdW) hexagonal boron nitride (*h*-BN) (bulk *h*-BN is non-polar).[10-12] Interestingly, the mechanism of the out-of-plane polarization and its switching by the in-plane vdW sliding is fundamentally different from the slight deformations of tightly bonded atoms in common 3D and 2D ferroelectrics.[12] The origin of the vertical polarization in the naturally non-polar bilayer boron nitride (BBN) is from the parallel-stacking arrangement of two *h*-BN sheets into a metastable AB or BA configuration (see **Figure 1**), which breaks the inversion symmetry and redistributes the charge between the pairs of fully-eclipsed B-N.[9-10] The vertical electrical polarization in one configuration, such as AB, can be switched by laterally sliding one layer over the other with a half lattice constant. Because of the weak interlayer vdW interaction, the calculated sliding barrier from AB to BA is only a few meV,[9] and a quite small voltage of 0.4 V can switch the polarization in the experiment.[10-11] Such polarization-sliding locking in BBN is enabled by the lateral rigidity of *h*-BN. Besides the mechanical rigidity, insulating 2D *h*-BN is also well-known for its chemically inert and thermal stability, which is often used as the protection for other 2D materials.[13-17] The superior mechanical, electrical, chemical and thermal properties make sliding ferroelectric BBN an ideal unit to be integrated into the current silicon-based electronic devices or employed in future new technologies, such as ultrathin non-volatile memories (FRAMs), flexible ferroelectric field-effect transistors (FFETs), wearable nanosensors, and ferroelectric tunnel junctions (FTJs).[8]

Ferroelectric tunnel junctions, composed of two metal electrodes separated by a thin ferroelectric barrier, can achieve different tunnelling electroresistance (TER) through polarization switching.[18-21] Applying the vdW sliding ferroelectric BBN as the ultrathin barrier can build 2D sliding ferroelectric tunnel junctions (SFTJs). Ferroelectric BBN has clean, uniform and weak interfacial coupling in its vdW structures as well as the polarization switching by an applied voltage. This paves a new way to realize low-power nanoscale tunnel junctions with novel functionalities.

In this work, we develop SFTJs using the sliding ferroelectric BBN as the tunnel barrier and evaluate the corresponding devices' transport behaviour by density functional theory (DFT) combined with the nonequilibrium Green's function (NEGF) method. We first study the ferroelectric transport of the BBN-based SFTJs by directly combining different stacked BBN with conventional Au electrodes. It is surprised to find that the tunnelling electroresistance is close to zero, indicating no ferroelectricity in BBN, which is in contrast to the experimental reports. We unveil such unexpected result by the study of metal contacts on ferroelectric BBN. The strong hybridization between Au and the adjacent *h*-BN diminishes the charge-reorganized ferroelectricity between BBN. We, thus, introduce graphene (Gr) intercalation layers between Au and *h*-BN as a protection to the BBN ferroelectric unit, which really generates an apparent ferroelectric output. According to the analysis of metal contacts and the comparison of interfacial properties of Au/*h*-BN and Au/Gr/*h*-BN, we point that graphene or other high-mobility vdW materials is a necessary protective component when building 2D vdW sliding ferroelectric tunnel junctions or other ferroelectric devices.

**Results**

Bilayer BN in AB and BA stacking structures possesses spontaneous polarization, which is oriented normal to the plane due to a net charge transfer induced by crystal asymmetry.[9] The AB and BA stacks can be obtained from layer-sliding the AA stack (**Figure 1a**). When the sliding layer of the AA stack slides $\frac{1}{2}a$ downwards the *x* axis ($a$ is the lattice parameter of BBN), AB stack is formed (**Figure 1b**). In this configuration, the boron atom in the sliding layer is overhead the nitrogen atom in the fixed layer, and the horizontal alignment of the $2p_z$ orbitals of N and B atoms distorts the orbital of N atoms, which gives rise to an out-of-plane electric dipole moment along the $-z$ axis shown in **Fig. 1b**.[9-11] If sliding $\frac{1}{2}a$ upwards the *x* axis, the BA stack is formed (**Fig. 1c**), where the nitrogen atoms in the sliding layer are overhead of the boron atoms in the fixed layer, and the opposite dipole moment along the $+z$ axis is generated.

The property of ferroelectricity of BBN can be applied for making ferroelectric devices. To investigate the sliding ferroelectric behaviour within the reported ferroelectric BBN, we use BBN in AA-, AB-, and BA-stack types contacted on Au (111) electrodes to build SFTJs (**Figs. 1d-1i**) and calculate their tunnelling electroresistance. To construct Au/BBN/Au junctions, we fix the in-plane primitive lattice constant of BBN to $a$=2.51 Å, build a 2×2 supercell, and adapt the lattice constant of Au (111) accordingly with the lattice mismatch of only 0.2%. The sectional views of Au/BBN in different stacking configurations are shown in **Figs. 1d-1f**. The optimized interlayer distances between two *h*-BN layers, and between Au and the neighboring *h*-BN flake are 3.10 Å and 3.37 Å, respectively, in agreement with previous results.[9, 22]

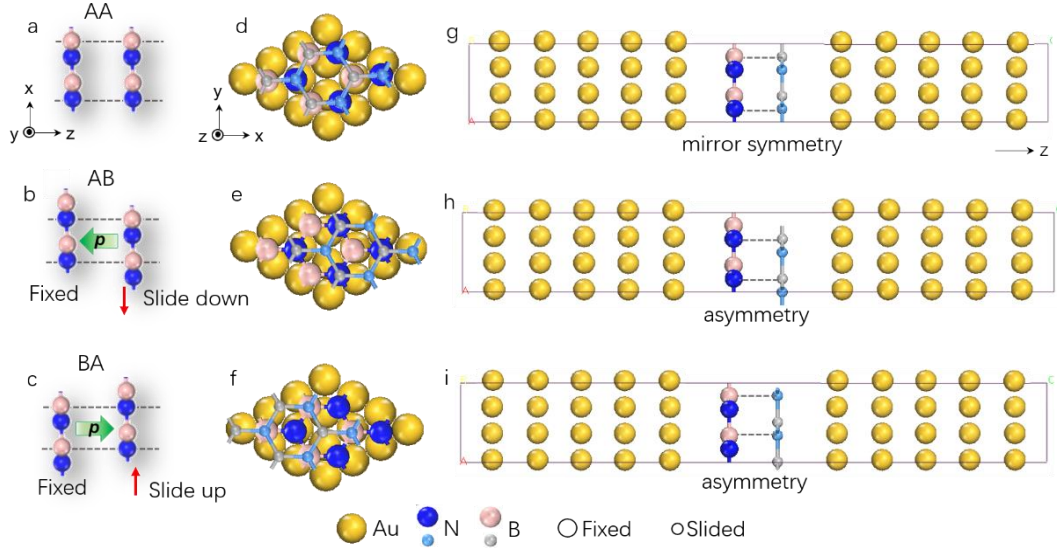

**Figure 1.** Schematic illustration of the atomic arrangement for AA- (a), AB- (b), and BA- (c) stacked bilayer *h*-BN. (d-f) Sectional view of bilayer *h*-BN with the different atomic configurations on Au (111). For clarity, atoms of the sliding layer *h*-BN are represented by small balls. (g-i) Side view of the device configuration of bilayer *h*-BN with different atomic arrangements sandwiched by two semi-infinite Au (111) electrodes.

It is known that in common FTJ devices, the opposite ferroelectric polarization states have different current responses when a bias voltage, $V_b$, is added to evaluate the transport behaviour.[10-11] Thus, it is naturally expected that the BBN-based FTJs should have the same phenomenon as common FTJs. Surprisingly, the I-V outputs of the two opposite ferroelectric states, AB and BA, are almost the same as shown in **Fig. 2a**. When $V_b$ sweeps from 0 to 1.2 V, both $I_{AB}$ and $I_{BA}$ linearly increase to ~600 μA/μm, and slightly higher than the current of non-polarized AA (382 μA/μm). We reveal such puzzle by the Mulliken population analysis. It is found that electrons are transferred from Au to *h*-BN, forming interfacial dipoles.[22] Each *h*-BN layer receives around 0.022*e* per primitive cell from the adjacent Au, and the net electric dipole is zero (AA) or close to zero (AB and BA) for the whole system (**Fig. 2b**). Such strong coupling between the Au electrode and *h*-BN results in a negligible mutual charge transfer within BBN itself. In such scenario, the spontaneous polarization in the intrinsic AB and BA stacks disappears, and the net dipole of AB and BA is almost equal to zero as nonpolar AA (**Fig. 2b**), resulting in the similar I-V curve of AB and BA as shown in **Fig. 2a**. It is worth noting that there is slight difference between $I_{AB}$ and $I_{BA}$ in **Fig. 2a** because of the existence of a tiny net dipole caused by a non-mirror polarization field in the AB and BA stacks.

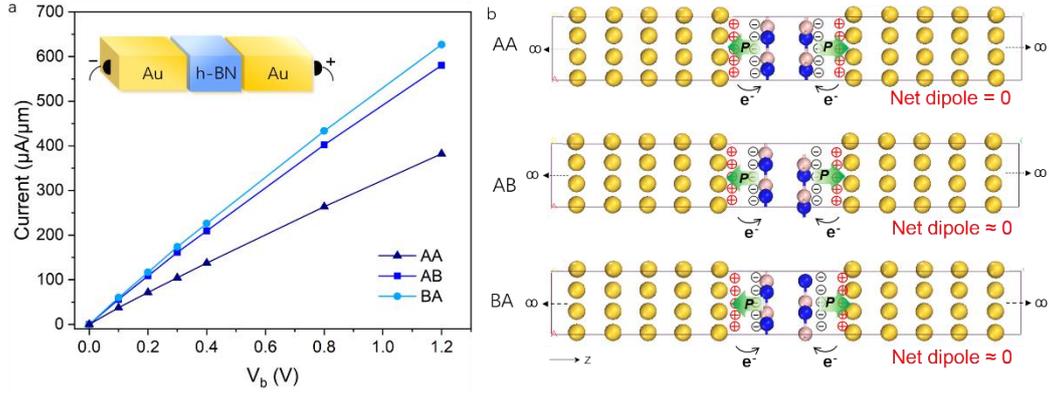

**Figure 2.** (a) I-V outputs of bilayer *h*-BN sliding ferroelectric tunnel junction (Au/BBN/Au) with different stacking configurations, AA, AB and BA. (b) Interfacial dipole diagram for the three *h*-BN configurations. The formation of the dipole is illustrated by the electron transfer. + and – are the positive and negative charges, respectively.

Once understanding the origin of the disappearance of ferroelectric polarization in BBN, we next figure out how to recover it in Au/BBN/Au SFTJs. In order to recover the ferroelectricity (or charge transfer) within two *h*-BN layers, a natural idea is to separate Au and BN to break their strong interaction. Breaking the interaction by introducing a vacuum spacing at the interface is not a practical way to make a metal-insulator contact. However, inserting a 2D vdW layer between the metal and BN can be. This layer has to be sufficiently thin as to not form a high barrier for electron transport. A monolayer of graphene is ideally suited. A single layer of graphene can be deposited or grown or transferred experimentally on Au substrates or BN.[10] It has been reported that depositing a monolayer graphene on Au substrates gives a metallic conduction,[10] which indicates that the graphene layer is transparent to electrons.

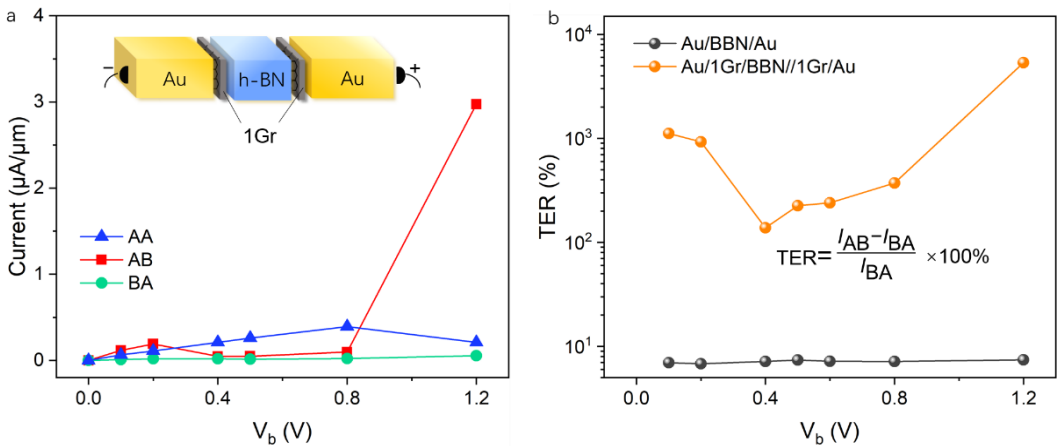

**Figure 3.** (a) I-V outputs of bilayer *h*-BN ferroelectric tunnel junctions with monolayer graphene intercalation (Au/1Gr/BBN/1Gr/Au). (b) The calculated tunnelling electroresistance of the bilayer *h*-BN sliding ferroelectric tunnel junction with and without monolayer graphene intercalation.

We first consider monolayer graphene (1Gr) and the layer-dependent effect is discussed later. A 2×2 supercell monolayer graphene is inserted between Au and *h*-BN in the previous Au/BBN/Au structures, forming Au/1Gr/BBN/1Gr/Au SFTJs (inset of **Fig. 3a**). The I-V curves of the new devices are shown in **Fig.3a**. It is firstly noted that $I_{AB}$ is two orders of magnitude less than that of the Au/BBN/Au FTJs under 1.2 eV. The reason lies in the thicker tunnel barrier with the graphene intercalation. Although the absolute current is decreased, the difference between $I_{AB}$ and $I_{BA}$ after inserting graphene is much greater than that without graphene, resulting a giant tunnelling electroresistance as shown in **Fig. 3b**. The TER of the Au/BBN/Au SFTJ keeps only ~5%, while it is up to 10,000% at 1.2 V in the Au/1Gr/BBN/1Gr/Au SFTJ. The slightly drop of the TER at 0.4 V might be due to the discontinuous states of monolayer graphene along the transport direction. Although the bias window gets wider from 0.1 to 0.4 eV, a very small $T(E)$ induced by the zero states of graphene produces a reduced current.[23] Once the energy range of the discontinuous states is fully included in or excluded from the bias window, $T(E)$ and the current will increase from 0.4 to 1.2 eV. In general, the TER increases up to 4 orders of magnitude with the resort of graphene.

On the basis of the large difference between $I_{AB}$ and $I_{BA}$, we believe the ferroelectricity in the AB- and BA-stack BBN is recovered by inserting graphene. To check this, we calculate the charge transfer between Au and the inserted graphene and their internal build-in polarization field. **Figure 4a** shows that the Au/Gr interfaces located at the two ends have the opposite polarization field, and the net dipole stems only from the BBN part. Therefore, the AA-stack FTJ has zero dipole moment, while the AB- and BA-stacks have the non-zero polarization and opposite in sense. The Mulliken population analysis shows that the interlayer voltage *U* is 0.27 V and 0.18 V in the AB- and BA-stacked bilayer h-BN in Au/1Gr/BBN/1Gr/Au, respectively, in good agreement with the reported results of the pristine BBN,[9] indicating the recovery of BBN ferroelectricity in Au/1Gr/BBN/1Gr/Au.

To further demonstrate the protective role of graphene in the SFTJs, the BN $p_z$ orbit projected band structures of pristine *h*-BN, Au/*h*-BN, and Au/1Gr/*h*-BN are shown in **Fig. 4b**. Compared with the pristine band dispersion of *h*-BN, the $p_z$ orbital bands of BN are highly interrupted by Au in the Au/*h*-BN heterostructure due to their strong coupling. After inserting monolayer graphene between Au and *h*-BN to screening the Au/BN interaction, the *h*-BN $p_z$ orbit becomes almost as continuous as the pristine *h*-BN. We also plot the planar averaged macroscopic electrostatic potential with and without graphene in **Fig. S1** to confirm the reservation of ferroelectric polarization in BBN by the graphene intercalation. To exclude the possible interfacial effects between Gr/*h*-BN, we do charge density difference calculations in two interfaces (Au/Gr and Gr/*h*-BN interface) (**Fig. S1**). It is found that no matter in the AB or BA configuration, electrons are locally distributed in the Au/Gr interface rather than the Gr/*h*-BN interface.

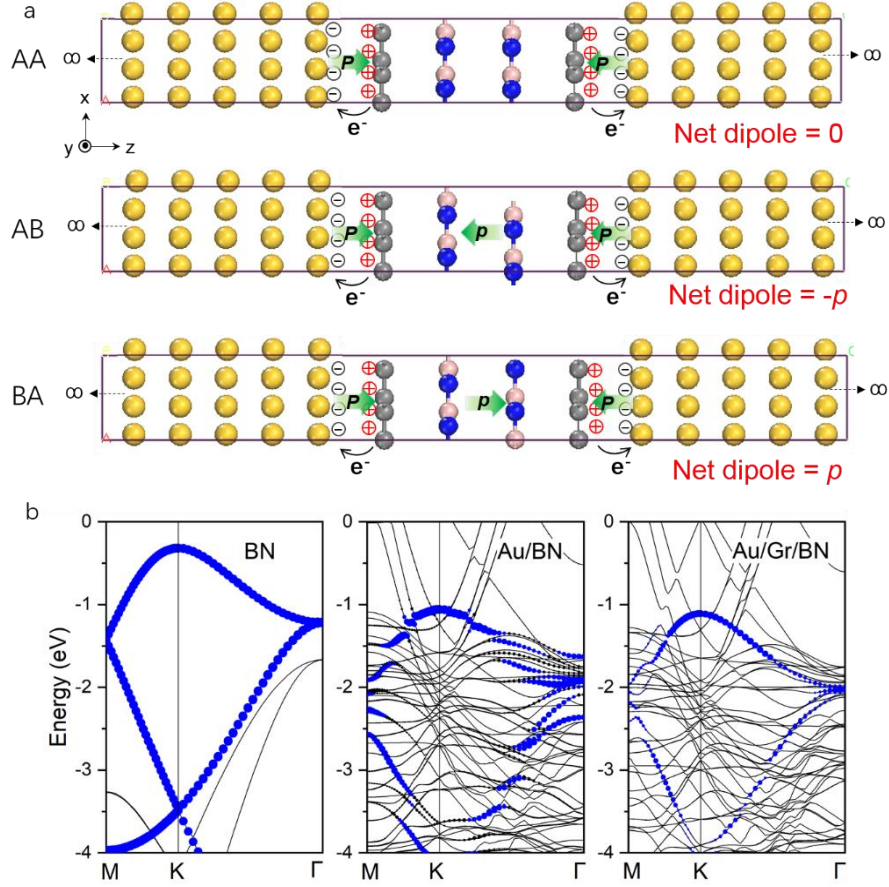

**Figure 4.** (a) The interface dipole diagram for the different BBN arrangements with monolayer graphene intercalation. The dipoles are illustrated by green arrows. (b) Band structures of monolayer *h*-BN in a 2×2 supercell, Au/1BN, and Au/1Gr/1BN. The blue points correspond to the *h*-BN $p_z$ orbital which is along the transport direction. The Fermi level is set to zero.

Finally, we investigate the metal-ferroelectrics contact. A schematic illustration of the band edges at the Au/*h*-BN interface with and without graphene intercalation is drawn in **Figs. 5a** and **5b**, and the detailed parameters can be found in **Table S1**. It is found that monolayer *h*-BN forms a p-type Schottky barrier ($\Phi_p^W$) of 1.12 eV and 1.10 eV in the interface of Au/*h*-BN and Au/Gr/*h*-BN, respectively.[22] As can be seen, the change of $\Phi_p^W$ is only 0.02 eV after inserting graphene at the interface. It is worth noting that at the Heyd Scuseria Ernzerhof (HSE) level, the difference is still 0.02 eV although the $\Phi_p^W$ becomes 1.90 eV and 1.88 eV for two cases, respectively.[22, 24] Therefore, the graphene intercalation almost does not change the Schottky barrier height.

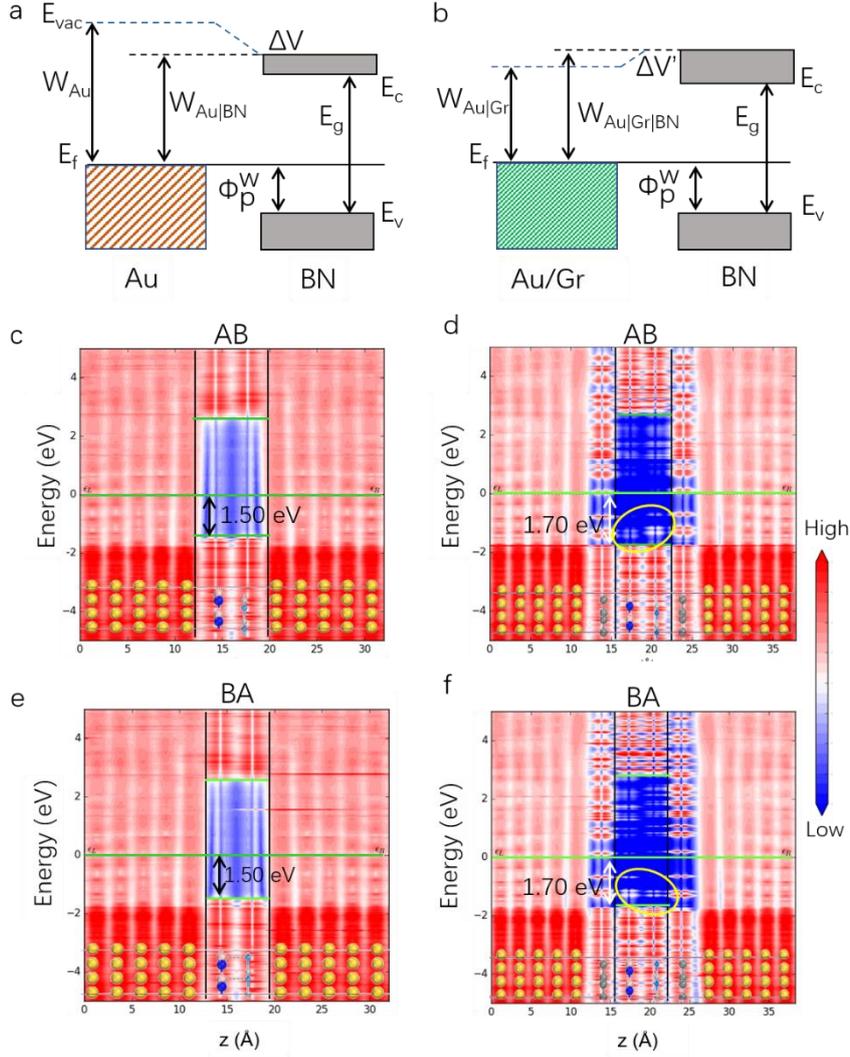

**Figure 5.** Schematic band alignment of the Au/*h*-BN interface (a) and the Au/1Gr/*h*-BN interface (b). $E_{vac}$ and $E_f$ are the vacuum level and the Fermi level, respectively. $E_c$, $E_v$, $E_g$ are the conduction band minimum, valence band maximum, and bandgap of *h*-BN, respectively. Position-resolved local density of states for the bilayer *h*-BN ferroelectric tunnel junction without (c) and with graphene intercalation (d). (e-f) The same as c) and d) but for the BA-stacked bilayer *h*-BN. The Fermi level is set to zero. The Schottky barriers are labelled by the double arrow. The yellow circles highlight the different states on two BN layers under the AB and BA configurations.

To more accurately describe the interfacial property from the device point of view, we calculate and analyse the local density of states (LDOS) of the SFTJs with and without the graphene intercalation through the quantum transport simulations. **Figures 5c** and **5e** show that the LDOS of the Au/BBN/Au FTJs in the AB and BA stacks. As can be seen, both have the same DOS on the two BN layers, which again proves the non-ferroelectric current response of the Au/BBN/Au devices. It is also found that there are many metal-induced states in the bandgap of BBN because of the strong coupling between Au and BN. The Schottky barrier height ($\Phi_p^{QT}$ from the quantum transport) formed at the Au/BBN interface is 1.50 eV, which is slightly higher than that ($\Phi_p^W$) obtained from the roughly work-function approximation. The difference between $\Phi_p^{QT}$ and $\Phi_p^W$ is mainly because the quantum transport takes the device environment into consideration rather than only the interface.[25] After inserting monolayer graphene, the big LDOS difference of the AB- and BA-stack SFTJs emerges in two BN layers (**Figs. 5d-5f**), labelled by yellow circles—the asymmetric states appear above the valence band maximum of BBN. In the AB-stacked SFTJ, more states are distributed on the right *h*-BN layer region than the left layer region. Whereas, the distribution is reversed in the BA-stacked one. This phenomenon supports their opposite polarization direction. Besides, $\Phi_p^{QT}$ becomes 1.70 eV, a bit larger than that without graphene intercalation, constant with that by the work-function analysis. Moreover, the metal-induced states in the bandgap of BBN get drastically decreased after inserting graphene and even absent nearby the Fermi level. It is also noteworthy that the graphene states (around 15 Å and 25 Å) are inconsecutive especially around the Fermi level, which is the main reason for the non-monotonic increasing current within the certain bias window in the transport calculations in **Fig. 3a**.

**Discussion**

Note that making any device with vdW ferroelectrics necessarily involves making contacts with metals. When building vdW vertical ferroelectric devices, ferroelectricity reservation has to be taken into consideration. In this work we propose the intercalation of high conductive monolayer graphene as a solution to preserve the ferroelectricity, which is also used as the charge carrier detector in the experiment.[10] To prove the generalization of our solution, we also study the TER by inserting other vdW layers, including bilayer graphene and monolayer *h*-BN (**Figure S3**). Based on our calculations, if only one additional *h*-BN layer is intercalated, the conductivity is comparable to that of monolayer graphene due to the strong interaction between Au and *h*-BN. This is in good agreement with the experiment, in which Yasuda *et al*. encapsulated ferroelectric BBN by *h*-BN in the experiment.[10] Although any vdW material can be used as the physical protection layer for sliding ferroelectrics, our calculations provide two suggestions: i) carrier transparent vdW materials are strongly recommended for keeping the high device conductivity if thick intercalation layers are used, such as multilayer 2D materials or monolayer 2D materials with multiple atomic layers; and ii) the inserted 2D materials should keep the similar Schottky barrier or better lower it.

For electrodes, although graphene is commonly used as the electrode due to its semi-metal nature, the transport conductivity with multilayer graphene electrodes is 4 orders of magnitude less than that of Au electrodes based on our calculations (**Table S2**). Gold, a well-used electrode material with a large work function, can provide a large number of electrons compared with multilayer graphene electrodes, resulting in high conductivity. Other metal electrodes forming a small Schottky barrier or even Ohmic contact with *h*-BN are also highly recommended for further improving the conducting behaviour.

Besides the sliding ferroelectricity of BBN, other bipartite honeycomb 2D materials, such as 2H-phase transition metal dichalcogenides (TMDs, like $MoS_2$ and $WSe_2$) and group III chalcogenides (GaS and GaSe), have been reported as the candidates of 2D sliding ferroelectricity. [9, 26-27] Very recently, Wang *et al.* experimentally reported $MoS_2$, $MoSe_2$, $WS_2$ and $WSe_2$-based sliding ferroelectrics.[28] Because of the wide bandgap of *h*-BN, the tunnelling current of BBN SFTJs is quite small unless a huge $V_b$ drives thermal current. The vdW SFTJs built by 2H-phase transition metal dichalcogenides or group III chalcogenides might be easier to have high current output because of their narrower bandgaps (around 2 eV). Ideally, once a small bias (~2 V) covering their bandgaps range is applied, the currents are drastically enhanced by thermal carriers, which pass from one lead to the other above or beneath the bandgap edges of the tunnel region. Furthermore, the robust ferroelectricity in bilayer TMDs enriches the properties and applications of devices by the interplay between ferroelectricity and the abundant physical properties of TMDs, such as ferroelectric-optoelectronics and ferroelectric-valleytronics.

**Conclusion**

Inspired by the recent observation of the sliding ferroelectricity in vdW bilayer *h*-BN, we develop sliding FTJs and evaluate the ferroelectric behaviour by the quantum transport calculations. The unexpected zero polarization in SFTJs results from the strong Au/*h*-BN interfacial electric field when directly deposit bilayer BN on Au electrodes. The Au/BN contact electric field disrupts the charge transfer within BBN, which is the origin of ferroelectric polarization. With the protection of graphene layers, the sliding ferroelectricity in Au/1Gr/BBN/1Gr/Au SFTJ is recovered. After comparing the interfacial properties with and without graphene intercalation from work function approximation and quantum transport results, we conclude that monolayer graphene would not change the other interfacial properties, such as the Schottky barrier. Different resistance states by stacking engineering in the vdW sliding FE-based tunnel junctions indicate the multiple non-volatile feature. The calculated ferroelectric output, giant TER and multiple non-volatile states of the BBN-based sliding FTJs allow us to recognize the potential ferroelectric applications of the layer-engineered 2D vdW sliding ferroelectrics.

**Methods**

The transport properties are calculated using the density functional theory coupled with the nonequilibrium Green's function formalism, implemented in the Atomistix ToolKit software package.[29-30] Following the Landauer-Bűttiker formula, the current under a given bias $V_b$ is calculated by the below formula:[23]

$$I(V_b)=\frac{e}{h}\int_{-\infty}^{+\infty}\{T(E,V_b)[f(E-\mu_L)-f(E-\mu_R)]\}dE,$$

where $T(E, V_b)$ is the transmission coefficient, $f$ is the Fermi-Dirac distribution function, and $\mu_{L/R}$ is the Fermi level of the left/right electrode. The transmission coefficient $T(E)$, which is the average overall different $k_{//}$ in the irreducible Brillouin zone, can be calculated by:[23]

$$T^{k_{//}}(E) = Tr[\Gamma_l^{k_{//}}(E)\, G^{k_{//}}(E)\, \Gamma_r^{k_{//}}(E)\, G^{k_{//}\dagger}(E)],$$

where $G^{k_{//}}(E)$ and $G^{k_{//}\dagger}(E)$ stand for the retard and advanced Green's function, respectively, and $\Gamma_{l/r}^{k_{//}}(E) = i(\sum_{l/r} - \sum_{l/r}^{\dagger})$ is the level broadening width stemming from left/right electrode in the form of self-energy $\sum_{l/r}$.

The periodic conditions are considered in the electrode region along the $x$, $y$, and $z$ directions. The period (Dirichlet) condition is applied to the central region along the $x$ and $y$ ($z$) directions of the BBN slice (**Figure 1**).

A generalized gradient approximation (GGA) in the form of Perdew-Burke-Ernzerhof (PBE) potential is adopted throughout the device calculations.[31-32] The double zeta polarized (DZP) set is used. The real-space mesh cutoff is taken as 155 Hartree. The $k$-point meshes for the electrode region and the central region are 8×8×107 and 8×8×1, respectively. The DFT-D3 correction is chosen to describe the vdW interaction.

The Schottky barrier height is evaluated by $\Phi_p^W = E_V - W_M + \Delta V$, where, $E_V$ is the valence band maximum of $h$-BN, $W_M$ is the work function of the Au or Au/Gr surface, and $\Delta V$ is the potential step formed at the contact interface with $\Delta V = W_M - W_{M/BN}$, where M = Au, Au/Gr.

**Acknowledgments**

We thank Dr. Weilong Kong for fruitful discussions. This work was supported by the National Key Research and Development Program of China (Nos. 2017YFA0206303 and 2016YFB0700600 (National Materials Genome Project)), the National Natural Science Foundation of China (Nos. 11674005, 91964101, 51731001, and 11975035), Singapore MOE Tier 1 (Grants No. R-265-000-651-114 and No. R-265-000-691-114), China Scholarship Council, High-Performance Computing Platform of Peking University, and High-Performance Computing (HPC) of National University of Singapore.

**Conflict of interest**

# TOC

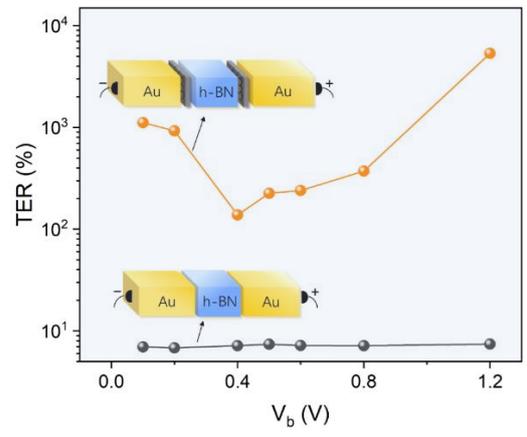